\documentclass[12pt,cite,epsf]{article}

\usepackage{graphicx}

\def\bea{\begin{eqnarray}}
\def\eea{\end{eqnarray}}

\def\a{\alpha}

\def\e{\epsilon}

\def\m{\mu}
\def\n{\nu}

\def\D{\Delta}
\def\F{\Phi}
\def\G{\Gamma}

\def\T{{\bf 10}}
\def\F{{\bf 5}}
\def\Fb{{\bf \bar{5}}}
\def\1{{\bf 1}}
\def\M{{\bf 16}}
\def\Mb{{\bf\bar{16}}}
\def\Mp{{M_{\rm P}}}
\def\mG{{m_{3/2}}}
\def\mt{\widetilde{m}}

\def\bo{{\raise.15ex\hbox{\large$\Box$}}}               
\def\face{{\raise.2ex\hbox{$\displaystyle \bigodot$}\mskip-2.2mu \llap {$\ddot
        \smile$}}}                                      
\def\dg{\dagger}                                     

\def\leftrightarrowfill{$\mathsurround=0pt \mathord\leftarrow \mkern-6mu
        \cleaders\hbox{$\mkern-2mu \mathord- \mkern-2mu$}\hfill
        \mkern-6mu \mathord\rightarrow$}       
\def\dvec#1{\vbox{\ialign{##\crcr
        \leftrightarrowfill\crcr\noalign{\kern-1pt\nointerlineskip}
        $\hfil\displaystyle{#1}\hfil$\crcr}}}           

\def\beq{\begin{equation}}
\def\eeq{\end{equation}}

\def\lsim{\mathrel{\raise.3ex\hbox{$<$\kern-.75em\lower1ex\hbox{$\sim$}}}}
\def\gsim{\mathrel{\raise.3ex\hbox{$>$\kern-.75em\lower1ex\hbox{$\sim$}}}}
\def\Rp{\not R_p}

\catcode`@=11
\def\@citex[#1]#2{\if@filesw\immediate\write\@auxout{\string\citation{#2}}\fi
  \def\@citea{}\@cite{\@for\@citeb:=#2\do
    {\@citea\def\@citea{,\penalty\@m}\@ifundefined
      {b@\@citeb}{{\bf ?}\@warning
       {Citation `\@citeb' on page \thepage \space undefined}}%
\hbox{\csname b@\@citeb\endcsname}}}{#1}}

\def\citer{\@ifnextchar [{\@tempswatrue\@citexr}{\@tempswafalse\@citexr[]}}
\def\@citexr[#1]#2{\if@filesw\immediate\write\@auxout{\string\citation{#2}}\fi
  \def\@citea{}\@cite{\@for\@citeb:=#2\do
    {\@citea\def\@citea{--\penalty\@m}\@ifundefined
       {b@\@citeb}{{\bf ?}\@warning
       {Citation `\@citeb' on page \thepage \space undefined}}%
\hbox{\csname b@\@citeb\endcsname}}}{#1}}
\catcode`@=12

\begin{document}
\date{\mbox{ }}

\title{ 
{\normalsize     
DESY 06-244\hfill\mbox{}\\
UT-07-03\hfill\mbox{}\\
February 2007\hfill\mbox{}\\}
\vspace{2cm}
\bf Gravitino Dark Matter\\ in R-Parity Breaking Vacua \\[8mm]}
%
\author{W.~Buchm\"uller$^a$, L.~Covi$^a$, K.~Hamaguchi$^b$, 
A.~Ibarra$^a$, T.~T.~Yanagida$^{b}$\\[2mm]
{\normalsize\it a Deutsches Elektronen-Synchrotron DESY, Hamburg, Germany}\\
{\normalsize\it b Department of Physics, University of Tokyo, Tokyo, Japan}}
\maketitle

\thispagestyle{empty}

\vspace{1cm}
\begin{abstract}
\noindent
We show that in the case of small R-parity and lepton number breaking 
couplings, primordial nucleosynthesis, thermal leptogenesis and
gravitino dark matter are naturally consistent for gravitino masses
$m_{3/2} \gsim 5~{\rm GeV}$. We present a model where R-parity 
breaking is tied to B-L breaking, which predicts the needed small 
couplings. The metastable next-to-lightest superparticle has a
decay length that is typically larger than a few centimeters, 
with characteristic signatures at the LHC. The photon flux 
produced by relic gravitino decays may be part of the apparent 
excess in the extragalactic diffuse gamma-ray flux obtained from 
the EGRET data for a gravitino mass $m_{3/2} \sim 10~{\rm GeV}$. 
In this case, a clear signal can be expected from GLAST in the near 
future.
\end{abstract}

\newpage

\section{Introduction}

Most supersymmetric extensions of the standard model impose R-parity 
\cite{fay78}
as an exact symmetry of the supergravity Lagrangian. In this way,
one forbids renormalizable baryon and lepton number violating interactions 
which might cause too rapid proton decay \cite{syw81}. 
On theoretical grounds, however,
theories with and without R-parity are on the same footing, and in
low-energy effective theories obtained from string compactifications
R-parity plays no preferred role.

One can also construct supersymmetric extensions of the standard model
without R-parity \cite{hs84}, and the phenomenological constraints on these
theories have been studied in great detail \cite{axx04}. Without R-parity
conservation, the lightest superparticle (LSP) is no longer stable and, 
in general, it does not contribute to dark matter.

Stringent constraints on the lepton number and R-parity violating interactions
\begin{equation}\label{deltaL}
W_{\D L=1} = \lambda_{ikj} l_i e^c_j l_k + \lambda'_{kji} d^c_i q_j l_k
\end{equation}
are imposed by baryogenesis. Both operators contain lepton doublets.
Together with sphaleron processes they therefore influence the baryon
asymmetry at high temperature in the early universe. The requirement that 
an existing baryon asymmetry is not erased before the electroweak transition 
typically implies \cite{cxx91}
\begin{equation}\label{rpv}
\lambda\ ,\lambda' < 10^{-7}\;.
\end{equation}
It is very remarkable that for such a small breaking of R-parity a gravitino
LSP has a lifetime much longer than the age of the universe \cite{ty00}.
This is due to the double suppression by the inverse Planck mass and the
R-parity breaking coupling, $\G_{3/2} \propto \lambda^2 m_{3/2}^3/\Mp^2$. 
We find for the gravitino lifetime
\begin{equation}
\tau_{3/2}\ \sim \ 10^{26} {\rm s}\  
\left(\frac{\lambda}{10^{-7}}\right)^{-2}
\left(\frac{m_{3/2}}{10~{\rm GeV}}\right)^{-3}\;,
\end{equation}
which is consistent with gravitino dark matter.

For a gravitino LSP, the properties of the next-to-lightest superparticle
(NLSP) are strongly constrained by primordial nucleosynthesis (BBN).
In the particularly interesting case of a charged NLSP, like a scalar
$\tau$-lepton, its lifetime has to be relatively short, 
$\tau_{\rm NLSP} \lsim 10^{3} - 10^{4}~{\rm s}$ \cite{pos06}\footnote{See also
\cite{kt06}. Here, we consider $m_{\rm NLSP} = {\cal O}(100~{\rm GeV})$. For
a heavier charged NLSP, $m_{\rm NLSP} > {\cal O}(1~{\rm TeV})$, the bound
on the lifetime becomes even more stringent (cf.~\cite{kkm05}).
We do not consider a late time entropy production in this paper, which is an another possible way to avoid these BBN constraints~\cite{BHIY}.}, 
which typically requires $m_{3/2} < 1~{\rm GeV}$.
Even for neutral particles, BBN excludes a neutralino NLSP  
for lifetimes longer than $ 10^{2}~{\rm s}$ due to
the strong constraints from hadronic showers \cite{kkm05}.
Only a sneutrino NLSP could be marginally acceptable also with longer
lifetimes, and therefore larger gravitino mass, in the region where the 
hadronic branching ratio of the decay is below $10^{-3}$ \cite{kkkm06}.

On the other hand, standard thermal leptogenesis \cite{fy86}, 
an attractive model 
for baryogenesis, needs a large reheating temperature in the early universe,
$T_R \gsim 10^9~{\rm GeV}$ (cf.~\cite{di02,bdp05}). This reheating 
temperature implies $m_{3/2} \gsim 5~{\rm GeV}$ for a gluino mass of 
$m_{\tilde{g}} = 500~{\rm GeV}$ in order to avoid overclosure of the
universe due to thermal gravitino 
production~\cite{bbb01,ps06}\footnote{We use the perturbative result 
for the gravitino production rate to leading order in the strong 
gauge coupling $g$. Since $g$ and also the thermal gluon mass are large,
the perturbative expansion is problematic \cite{bbb01}. 
The uncertainty due to higher orders in $g$ and nonperturbative effects 
is ${\cal O}(1)$. Possible effects due to thermal masses are also 
${\cal O}(1)$ \cite{rs07}.}. The lower
bound on the gravitino mass scales as 
$m_{3/2}^{\rm min} \sim T_R m_{\tilde{g}}^2$. 

All these cosmological problems are automatically solved without any fine
tuning of parameters in the case of a small breaking of R-parity, as given
in Eq.~(\ref{rpv}), with a gravitino LSP. The NLSP lifetime becomes 
sufficiently short for $\lambda, \lambda' > 10^{-14}$,
\begin{equation}
\tau_{\rm NLSP} \simeq 10^3 {\rm s} \left({\lambda\over 10^{-14}}\right)^{-2}
\left(m_{\rm NLSP}\over {100~{\rm GeV}}\right)^{-1}\;.
\end{equation}
Therefore,  primordial nucleosynthesis, thermal leptogenesis and gravitino 
dark matter are naturally consistent 
for $10^{-14} < \lambda,\lambda' < 10^{-7}$ and $m_{3/2} \gsim 5~{\rm GeV}$. 
This is the main point of this paper.

The paper is organised as follows. In Sec.~2 we present a model where R-parity 
breaking is tied to B-L breaking, yielding the needed small R-parity
breaking couplings. Sec.~3 deals with constraints from neutrino masses. 
Sec.~4 deals with implications for cosmology and collider physics. 
The results are discussed in Sec.~5.

\section{R-Parity Breaking and B-L Breaking}

\subsection{A Model of R-Parity Breaking} 

We consider a supersymmetric extension of the standard model whose symmetry
group $G$ includes $U(1)_{B-L}$ and R-invariance,
\begin{equation}
G = SU(3)\times SU(2)\times U(1)_Y\times U(1)_{B-L}\times U(1)_R\;.
\end{equation}
Three quark-lepton generations can be grouped into the $SU(5)$ representations
$\T_i = (q,u^c,e^c)_i$, $\Fb_i = (d^c,l)_i$ and $\1 = \n^c_i$, which
together form 16-plets of $SO(10)$. In addition, we have two Higgs doublets
$H_u$ and $H_d$, two standard model singlets $N^c$ and $N$, and three $SO(10)$ 
singlets $X$, $\Phi$ and $Z$. 
The two Higgs doublets are contained in $\F$- and $\Fb$-plets of $SU(5)$, 
which we shall also denote as $H_u$ and $H_d$, respectively. $N^c$ and $N$ are 
contained in $\M$ and $\Mb$ of $SO(10)$, which fixes their B-L charge 
to be $+1$ and $-1$, respectively. 
$X$, $\Phi$ and $Z$ have B-L charge zero. This set of 
fields is familiar from $SO(10)$ orbifold GUTs 
(cf.~\cite{abc03}): matter fields form complete $SO(10)$ 
representations, whereas fields which break $SU(2)\times U(1)_Y$ and 
$U(1)_{B-L}$ appear as `split' multiplets. For simplicity, we shall use
in the following often $SU(5)$ notation also for the Higgs multiplets.

The matter sector of the superpotential has the usual form
\begin{equation}\label{yuk}
W_M = h_{ij}^{(u)} \T_i\T_j H_u +  h_{ij}^{(d)} \Fb_i\T_j H_d
      + h_{ij}^{(\n)} \Fb_i \1_j H_d 
      +  {1\over M_{\rm P}} h_{ij}^{(n)} \1_i\1_j N^2\;,
\end{equation}
where $M_{\rm P} = 2.4 \times 10^{18}\ {\rm GeV}$ is the Planck mass.
The expectation values of the Higgs multiplets $H_u$ and $H_d$ generate 
Dirac masses of quarks and leptons, whereas the expectation value of the 
singlet Higgs field $N$ generates the Majorana mass matrix of the right-handed 
neutrinos $\1_i$. The superpotential responsible for B-L breaking is chosen as
\begin{equation}\label{bl}
W_{B-L} = X ( NN^c - \Phi^2)\; ,
\end{equation}
where unknown Yukawa couplings have been set equal to one.
$\Phi$ plays the role of a spectator field, which will finally be replaced
by its expectation value, $\langle \Phi \rangle = v_{B-L}$. Similarly,
$Z$ is a spectator field~\footnote{For simplicity, we use a 
spectator chiral superfield to describe supersymmetry breaking. 
The field $Z$ is not essential for the connection between 
R-parity breaking and B-L breaking discussed in this section.}, 
which breaks supersymmetry and $U(1)_R$, 
$\langle Z \rangle = F_Z \theta\theta$. The superpotential
in Eqs.~(\ref{yuk}) and (\ref{bl}) is the most general one consistent
with the \mbox{R-charges} listed in Table~1, 
up to higher order terms which we will discuss later. 
Note that the choice of a negative 
R-charge for $N^c$ forbids the dangerous superpotential terms
\begin{equation}
\Fb_i H_d N^c\;, \quad \Fb_i\Fb_j\T_k N^c\;,
\end{equation}
which would yield too large bilinear mixings and too rapid proton decay,
respectively.

The expectation value of $\Phi$ leads to the breaking of $B-L$,
\begin{equation}\label{breaking}
\langle N \rangle = \langle N^c \rangle = \langle \Phi \rangle = v_{B-L}\;,
\end{equation}
where the first equality is a consequence of the $U(1)_{B-L}$ D-term.
This generates a Majorana mass matrix $M$ for the  
right-handed neutrinos with three large eigenvalues, with $M_1 < M_2 < M_3$. 
If the largest eigenvalue of $h^{(n)}$ is ${\cal O}(1)$, one has 
$M_3 \simeq v_{B-L}^2/\Mp$. The heavy Majorana neutrinos can be integrated 
out yielding for the matter part of the superpotential
\begin{equation}
W_M = h_{ij}^{(u)}\ \T_i\T_j H_u +  h_{ij}^{(d)}\ \Fb_i\T_j H_d 
      - {1\over 2} (h^{(\n)}{1\over M}h^{(\n)T})_{ij}
        (\Fb_i H_u) (\Fb_j H_u) \;,
\end{equation}
with the familiar dimension-5 seesaw operator for light neutrino masses.

\begin{table}[t]
\begin{center}
\begin{tabular}{c|cccccccccc}\hline \hline
$ $ & $\T_i$ & $\F^*_i$ & $\1_i$ & $H_u$ & $H_d$ & $N$ & $N^c$ & 
$\Phi$ & $X$ & $Z$  \\ \hline
$R$ & 1 & 1 & 1 & 0 & 0 & 0 & -2 & -1 & 4 & 0 \\ \hline\hline
\end{tabular}
\medskip
\caption[dum]{\it R-charges of matter fields, Higgs fields and $SO(10)$
singlets.}
\end{center}
\end{table}

Since the field $\Phi$ carries R-charge $-1$, the VEV $\langle \Phi \rangle$
breaks R-parity, which is conserved by the VEV $\langle Z \rangle$.
Thus, the breaking of B-L is tied
to the breaking of R-parity. This is the key feature of the mechanism for
R-parity breaking presented in this paper\footnote{For a recent
discussion of the connection between B-L breaking and R-parity breaking
in the context of string compactifications, see \cite{tw06,bhx06}.}. 
The breaking of R-parity is transmitted to the low-energy degrees of freedom 
via higher-dimensional operators in the superpotential and the K\"ahler 
potential. The leading correction to the K\"ahler potential is
\begin{equation}
\delta K_1 =  {1\over \Mp^3}(a_i Z^\dg + a_i' Z)\Phi^\dg N^c \Fb_i H_u 
   + {1\over \Mp^3}(c_i Z^\dg + c_i' Z)\Phi N^\dg \Fb_i H_u + h.c. \;.
\end{equation}
Replacing the spectator fields $Z$ and $\Phi$, as well as $N^c$ by their 
expectation values, one obtains the correction to the superpotential
\begin{equation}\label{bilinear}
\delta W_1 =  \m_i \Theta \Fb_i H_u \;,
\end{equation}
with
\begin{equation}
\m_i = {\cal O}(\mG)\;, \quad 
\Theta = {v_{B-L}^2\over \Mp^2} \simeq {M_3\over \Mp}\;,
\label{eq:Theta}
\end{equation}
where $\mG = F_Z/(\sqrt{3}\Mp)$ is the gravitino mass. Note that $\Theta$
can be increased or decreased by an appropriate choice of Yukawa couplings
in Eqs.~(\ref{yuk}) and (\ref{bl}). Eq.~(\ref{bilinear})
is the familiar bilinear R-parity breaking term \cite{hs84}. 
The correction to the K\"ahler potential
\begin{equation}
\delta K_0 =  {k\over \Mp} Z^\dg H_d H_u + h.c.
\end{equation}
yields the corresponding R-parity conserving term \cite{gm88}
\begin{equation}\label{mu}
\delta W_0 =  \m H_d H_u \;, \quad \m = {\cal O}(\mG)\;.
\end{equation}
Note that $\m$ and $\m_i$ are generated by operators of different mass
dimension. Hence, their values may easily differ by one or two orders of
magnitude, allowing for $\mu > \mu_i, m_{3/2}$ and a gravitino LSP.

To analyse the complete superpotential including the R-symmetry breaking
terms, it is convenient to perform a rotation of the Higgs and lepton
superfields,
\begin{equation}
\label{rotation}
H_d = H_d' - \e_i l'_i\;, \quad l_i = l'_i + \e_i H_d'\;,
\end{equation}
where $\e_i = \m_i \Theta/\m$.
In terms of the new fields the superpotential reads
\begin{eqnarray}\label{pheno}
W &=& W_M + \delta W_0 + \delta W_1 \nonumber\\
&=& \mu H_d'H_u + h_{ij}^{(u)} q_i u^c_j H_u + h_{ij}^{(d)} d^c_i q_j H_d'
+ h_{ij}^{(e)} l'_i e^c_j H_d' \\
&& - \e_k h_{ij}^{(d)} d^c_i q_j l_k' - \e_k h_{ij}^{(e)}\ l'_i e^c_j l_k' 
- {1\over 2} (h^{(\n)}{1\over M}h^{(\n)T})_{ij}
        (l_i' H_u) (l_j' H_u) + {\cal O}(\e^2,\e m_\n) \nonumber\;.
\end{eqnarray}
The mixing of Higgs and lepton superfields has induced trilinear R-parity
breaking terms ${\cal O}(\e)$.  As we will discuss
in Sec.~3, the mixing terms induce vacuum expectation values
for the scalar neutrinos that in turn induce mixing
terms ${\cal O}(\e)$ of neutrinos with the neutralinos,
and neutrino masses suppressed by ${\cal O}(\e^2)$.

It is remarkable that the potentially dangerous operator leading to
proton decay is strongly suppressed compared to the trilinear terms 
${\cal O}(\e)$ in Eq.~(\ref{pheno}). The leading operator is
\begin{equation}
\delta W_2 = {1\over \Mp^5} u^c d^c d^c N^c \Phi^3 X\;.
\end{equation} 
For global supersymmetry one has $\langle X \rangle = 0$, which in 
supergravity is modified to $\langle X \rangle = {\cal O}(\mG)$~\footnote{The
VEV $\langle X \rangle = {\cal O}(\mG)$ also causes an additional
contribution to the bilinear term via $\delta W = 
(1/\Mp^3) X \Phi N^c \Fb_i H_u$, which is comparable to those from 
$\delta K_1$.}. One then obtains
\begin{equation}
\label{p-decay}
\delta W_2 \propto {\mG v_{B-L}^4 \over \Mp^5} u^c d^c d^c + \ldots \;.
\end{equation} 
For $\lambda,\lambda'$ satisfying Eq.~(\ref{rpv}),
the coefficient of the dangerous dimension-4 $\Delta B = 1$ operator
is much smaller than the upper bound from the proton lifetime \cite{axx04}.

\subsection{Scale of B-L breaking and Thermal Leptogenesis}

The phenomenological viability of the model depends on the size of 
R-parity breaking mixings $\e_i$ and therefore on the scale $v_{B-L}$ 
of R-parity breaking. An important constraint comes from baryogenesis.
As already discussed in the introduction, the potential washout of a 
baryon asymmetry before the electroweak phase transition is avoided if 
the R-parity violating Yukawa couplings satisfy 
$\lambda_{ijk},\lambda'_{ijk}\lsim 10^{-7}$,
which in turn implies:
\begin{equation}
\label{washout}
\left(\frac{\epsilon_i}{10^{-6}}\right)
\left(\frac{\tan\beta}{10}\right)\lsim 1\;.
\end{equation}
This is a sufficient
condition, which can be relaxed for some flavour structures \cite{cxx91}.

As an illustration for possible scales of B-L breaking we use a model
\cite{by99} for quark and lepton mass hierarchies based on a 
Froggatt-Nielsen
$U(1)$ flavour symmetry. The mass hierarchy is generated by the expectation 
value of a singlet field $\phi$ with charge $Q_{\phi}=-1$ via 
nonrenormalizable interactions with a scale 
$\Lambda = \langle \phi\rangle/\eta > \Lambda_{GUT}$, $\eta \simeq 0.06$.
Yukawa couplings and bilinear terms for $SU(5)$ multiplets $\psi_i$ with 
charge $Q_i$ scale like
\begin{equation}
h_{ij} \propto \eta^{Q_i + Q_j}\;, \quad \mu_i \propto \eta^{Q_i}\;.
\end{equation}
Charges $Q_i$ describing qualitatively the observed quark and lepton
masses and mixings are listed in Table~2. The model also predicts the
observed baryon asymmetry via leptogenesis for the cases where
$a+d=2$. There are two, at low energies indistinguishable, consistent 
scales of B-L breaking: $M_3 \sim 10^{15}~{\rm GeV}$ ($a=b=0$, $c=1$,
$d=2$) and $M_3 \sim 10^{12}~{\rm GeV}$ ($b=c=0$, $a=d=1$). For 
$\m_i/\m = 1.0 \ldots 0.01$ these two cases lead to the R-parity breaking 
mixing parameters (cf. Eq.~(\ref{eq:Theta}))
\begin{equation}
({\rm I})\quad {\e_i\over \eta^{Q_i}} = 10^{-3}\ldots 10^{-5}\;, \qquad
({\rm II})\quad {\e_i\over \eta^{Q_i}} = 10^{-6}\ldots 10^{-8}\; .
\end{equation}
In the extreme case $M_3 \sim M_2 \sim M_1 \sim 10^{10}~{\rm GeV}$
without Froggatt-Nielsen symmetry, where
leptogenesis may still work for an appropriate enhancement of the 
CP asymmetry, one has
\begin{equation}
({\rm III})\quad \e_i = 10^{-8}\ldots 10^{-10}\;. 
\end{equation}
In the flavour models (I) and (II) the RPV mixings $\e_{i}$ 
are suppressed by $\eta^{Q_i}$. As we shall see in the following section,
model (I) is inconsistent with the constraints from neutrino masses and
baryogenesis washout; the models (II) and (III) are consistent with both
constraints.

\begin{table}[t]
\begin{center}
\begin{tabular}{c|cccccccccccccc}\hline \hline
$\psi_i$ & $\T_3$ & $\T_2$ & $\T_1$ & $\F^*_3$ & $\F^*_2$ & $\F^*_1$ &
$\1_3$ & $\1_2$ & $\1_1$ & $H_u$ & $H_d$ & $\Phi$ & $X$ & $Z$ \\ \hline
$Q_i$ & 0 & 1 & 2 & a & a & a+1 & b & c & d & 0 & 0 & 0 & 0 & 0 \\ \hline\hline
\end{tabular}
\medskip
\caption{\it Chiral charges: $a=0$ or $1$, and $0\leq b \leq c \leq d$.} 
\end{center}
\end{table}

The expected mass scale of right-handed neutrinos depends on the mechanism
which breaks B-L. The expectation value of a field with lepton number
$L=2$ can generate heavy Majorana masses via renormalizable Yukawa
couplings. With B-L broken at the GUT scale, and for Yukawa coupling 
${\cal O}(1)$ for the third family, one then obtains the canonical result
$M_3 \sim v_{B-L} \sim 10^{15}~{\rm GeV}$. On the other hand, if right-handed 
neutrino masses are generated via a nonrenormalizable dimension-5 operator  
and the expectation value of a field with $L=1$, as in Eq.~(\ref{yuk}),
one has instead $M_3 \sim v_{B-L}^2/\Mp \sim 10^{12}~{\rm GeV}$. This
illustrates how the two mass scales for $M_3$, which correspond to the
two cases (I) and (II), respectively, might be obtained.

\section{Neutrino Masses}

The model we are considering generates after
supersymmetry breaking bilinear R-parity violating terms,
Eq.~({\ref{bilinear}), and tiny R-parity violating
Yukawa couplings, Eq.~(\ref{p-decay}), that we neglect in what follows.
Scenarios with just bilinear R-parity violation have 
been thoroughly studied in the literature \cite{bilinear}. 
Here, we will limit ourselves to estimate the size
of neutrino masses, following closely  \cite{ty99}. 

At the high-energy scale, the soft SUSY breaking Lagrangian
reads
\begin{equation}
\label{Lsoft-before}
-{\cal L}_{soft}= m^2_{H_d} |H_d|^2+m^2_{H_u} |H_u|^2+
m^2_{l_i} |{\tilde l_i}|^2+
(B H_d H_u + B_i \tilde l_i H_u + m^2_{l_i H_d}{\tilde l_i} H_d^*+h.c.)+...
\end{equation}
For the computation of neutrino masses we find convenient
to work in the basis where the R-parity violating bilinear couplings
in the superpotential are rotated away, $\mu_i=0$, through
the field redefinition Eq.~(\ref{rotation}). This choice of basis has the 
advantage that once the basis has been fixed at the
high energy scale, the condition $\mu_i=0$ holds at any 
scale, and it is not necessary to redefine the basis again
at low energies. We also choose the phases of the lepton
doublets such that the $\epsilon_i$ are real. In this
basis the soft SUSY breaking Lagrangian is given by
\begin{equation}
\label{Lsoft-after}
-{\cal L}_{soft}= m^2_{H'_d} |H'_d|^2+m^2_{H_u} |H_u|^2+
m^2_{l_i'} |{\tilde l'_i}|^2+
(B' H'_d H_u + B_i' \tilde l_i' H_u + m^2_{l_i' H'_d}{\tilde l_i'} 
{H'_d}^*+h.c.)+ ...
\end{equation}
where
\begin{eqnarray}
&m^2_{H'_d}=m^2_{H_d}+\epsilon_i{\rm Re}( m^2_{l_i H_d})
+{\cal O}(|\epsilon_i|^2)\;,\nonumber \\ 
&m^2_{l'_i}=m^2_{l_i}-\epsilon_i{\rm Re}(m^2_{l_i H_d})
+{\cal O}(|\epsilon_i|^2)\;,\nonumber \\
&B'=B+ B_i \epsilon_i\;,\nonumber\\
&B'_i=B_i-B\epsilon_i\;, \nonumber\\
&m^2_{l'_i H'_d}=m^2_{l_i H_d}+\epsilon_i (m^2_{l_i}-m^2_{H_d}) 
 +{\cal O}(|\epsilon_i|^2)\;.
\end{eqnarray}
Minimisation of the scalar potential yields non-vanishing
vacuum expectation values for the neutral components of the
Higgs doublets, as well as for the sneutrinos,
\begin{equation}
\label{snu-vev}
\langle \tilde\nu'_i\rangle=
\frac{B'_i\tan\beta+m^2_{l'_i H'_d}}
{m^2_{l'_i}-\frac{1}{2} M^2_Z \cos2\beta}
\langle H'_d\rangle\;, 
\quad v^2_\nu = \sum_{i=1}^3 \langle \tilde\nu'_i\rangle^2\;.
\end{equation}
These vacuum expectation values induce mixings between
neutrinos and gauginos, giving rise
to one non-vanishing neutrino mass through an
``electroweak'' seesaw mechanism. The resulting
neutrino mass is
\begin{equation}
\label{nu-mass}
m^{\Rp}_{\nu}=\frac{1}{2} g^2_Z v^2_\nu
\sum^4_{\a=1} {|c_{\tilde{z}\a}|^2\over m_{\chi^0_\a}}\;,
\end{equation}
where $g_Z$ is the Z-boson gauge coupling,
$m_{\chi^0_\a}$ are the neutralino masses
and $c_{\tilde{z}\a}$ is the Zino component of the neutralino $\chi^0_\a$.
In the following estimates we will replace
the sum over inverse neutralino masses by the inverse of the 
characteristic SUSY breaking scale, $1/\mt$.

The size of the neutrino masses depends crucially on the
mechanism for supersymmetry breaking. Generically, one expects
\begin{equation}
B'_i\sim \mu_i \mt \sim \epsilon_i \mt^2, 
\quad m^2_{l'_i H'_d}=\epsilon_i\mt^2\;.
\end{equation}
Then, using Eqs.~(\ref{snu-vev}) and (\ref{nu-mass}) we find
for $\e_1, \e_2 \leq \e_3$,
\begin{equation}
m^{\Rp}_\nu\sim 10^{-4} {\rm eV}
\left(\frac{\epsilon_3}{10^{-7}}\right)^2
\left(\frac{\mt}{200{\rm GeV}}\right)^{-1}\;.
 \label{mnurp}
\end{equation}

We can now insert the values of $\e_3$ for the different flavour models
considered in the previous section. Clearly, model (I), which has a 
high scale of B-L breaking, is excluded. 
In models (II) and (III) one has 
$\e_3 < 10^{-7}$ and $\e_3 < 10^{-8}$, respectively. Here the neutrino
mass terms induced by R-parity breaking are negligible and the 
baryogenesis constraint, Eq.~(\ref{washout}), is fulfilled for 
all values of $\tan\beta$.

\section{Cosmology and Collider Physics}

\subsection{Gravitino Decay}

Since R-parity is broken, the gravitino is no more stable in our
setting, but it still has a lifetime much longer than the age of 
the universe,
since it is suppressed both by the Planck mass and the small R-parity 
breaking parameters.

The two-body decay is determined by the mixing of the neutralinos 
with the
neutrinos. Neglecting the small neutrino masses, one has \cite{ty00}
\begin{eqnarray}
\label{gamma-gravitino}
\Gamma (\psi_{3/2} \rightarrow \gamma \nu)
= {1\over 32 \pi} |U_{\tilde\gamma\nu}|^2 \frac{m_{3/2}^3}{\Mp^2}\; .
\end{eqnarray}
The photino-neutrino mixing can be approximated by (cf.~Eq.~(\ref{snu-vev}))
\begin{equation}
\label{Ugammanu}
|U_{\tilde\gamma\nu}|  
\simeq g_z \left|\sum^4_{\a=1} c_{\tilde{\gamma}\a} c^*_{\tilde{z}\a}
{v_\nu \over m_{\chi^0_\a}}\right|
\sim 10^{-8} \left({\e_3\over 10^{-7}}\right)
\left({\mt\over 200~{\rm GeV}}\right)^{-1}\;,
\end{equation}
for $\e_1, \e_2 \leq \e_3$,
where we made the rough estimate $0.1/\mt$ for the weighted sum of
neutralino masses and the coupling, taking into account that not 
all mixings can be maximal.
Using $\Mp = 2.4 \times 10^{18} $ GeV, one obtains for
the gravitino lifetime \cite{ty00}
\begin{eqnarray}
\label{grav-lifetime}
\tau^{\rm 2-body}_{3/2} \simeq  4\times 10^{27} {\rm s} 
\left(\frac{\epsilon_3}{10^{-7}}\right)^{-2}
\left(\frac{\mt}{200~{\rm GeV}}\right)^{2}
\left(\frac{m_{3/2}}{10 \mbox{ GeV}} \right)^{-3}\; .
\end{eqnarray}

The three-body decay is usually subdominant due to the phase-space
and intermediate heavy particle suppression.
For the decay with intermediate heavy $\tilde\tau_R$, neglecting
all external masses in the phase space factor, we find
\begin{eqnarray}
\Gamma (\psi_{3/2} \rightarrow \tau_R l_i l_j)
= {|\lambda_{ij3}|^2 \over 3 (32)^2 \pi^3} 
\frac{m_{3/2}^3}{ M_P^2} 
F\left(\frac{m_{\tilde\tau_R}}{m_{3/2}} \right)\; ,
\end{eqnarray}
where 
\begin{equation}
F(\alpha ) = \int_0^1 dx \frac{x^4 (1-x)}{(1-x-\alpha^2)^2}
\simeq \frac{1}{30\alpha^4}\;.
\end{equation}
The full expression has been obtained in \cite{mc01}.

In the case where only bilinear R-parity breaking is present,
the $\lambda_{ijk}$ couplings are generated from the Yukawa couplings
as
\begin{equation}
\lambda_{ijk} = \epsilon_i h^{(e)}_{jk} \;.
\end{equation}
Then the inverse partial width for the three-body decay,
\begin{equation}
\Gamma (\psi_{3/2} \rightarrow \tau_R l_i l_j)^{-1} \simeq
2 \times 10^{37}s
\left({\e_2\over 10^{-7}}\right)^{-2}
\left(\frac{\tan\beta}{10}\right)^{-2}
\left( \frac{\mt}{200~{\rm GeV}} \right)^{4}
\left( \frac{m_{3/2}}{10~{\rm GeV}} \right)^{-7}\; ,
\end{equation}
is much larger than the lifetime determined from the two-body decay,
Eq.~(\ref{grav-lifetime}), as long as the mixing between photino
and neutrino is not unnaturally suppressed.

\subsection{Extragalactic Diffuse Gamma-Ray Emission}

A stringent astrophysical constraint for decaying gravitino dark matter
is the measured gamma-ray flux \cite{ty00}.
Assuming that the gravitino constitutes the dominant component of dark 
matter, its decay into neutrino and photon gives rise to an 
extragalactic diffuse gamma-ray flux 
with a characteristic energy spectrum, corresponding to a red shifted 
monochromatic line. A photon with measured energy 
\mbox{$E = m_{3/2}/(2(1+z))$}
has been emitted at the comoving distance $\chi(z)$, with
\mbox{$d\chi/dz=(1+z)^{-3/2}/(a_0 H_0\sqrt{\Omega_M(1+ \kappa (1+z)^{-3})})$}. 
Here $a_0$ and $H_0$ are the present scale factor and Hubble parameter, 
respectively, and $\kappa = \Omega_\Lambda/\Omega_M \simeq 3$, with 
$\Omega_\Lambda + \Omega_M =1$, assuming a flat universe.
For the photon flux one obtains, for $\tau_{3/2}\gg H_0^{-1}$,
\begin{equation}
\label{photon-flux}\label{signal}
 E^2 {dJ_{eg}\over dE} = C_\gamma 
\left(1+ \kappa \left({2E\over m_{3/2}}\right)^3\right)^{-1/2} 
\left({2E\over m_{3/2}}\right)^{5/2} 
\theta\left(1 - \frac{2E}{m_{3/2}}\right)\; , 
\end{equation}
with 
\begin{equation}
C_\gamma = \frac{ \Omega_{3/2} \rho_c}{8\pi\tau_{3/2}H_0 \Omega_M^{1/2}}
= 10^{-7}\ ({\rm cm}^2 {\rm str}~{\rm s})^{-1} {\rm GeV}
\left(\frac{\tau_{3/2}}{ 10^{28} \mbox{s}} \right)^{-1}\;;
\end{equation}
here $ \tau_{3/2} $ is given by Eq.~(\ref{grav-lifetime}),
and we have taken the gravitino density equal to the Cold Dark 
Matter density as $\Omega_{3/2} h^2 =0.1$, 
$\rho_c = 1.05\; h^2 \times 10^{-5} {\rm GeV} {\rm cm}^{-3}$,
$\Omega_M = 0.25$ and
$H_0 = h\; 100\; {\rm km}\; {\rm s}^{-1}\; {\rm Mpc}^{-1}$ 
with $h=0.73$~\cite{PDG}. 

In addition to the extragalactic signal one also expects a sharp line 
from the halo of our galaxy with an intensity comparable to the 
extragalactic signal and strong anisotropy~\cite{Asaka:1998ju}.
We have in fact from the decay of halo gravitinos
\begin{equation}
\label{photon-flux-halo}
 E^2 {dJ_{halo}\over dE} = D_\gamma\; 
\delta\left(1 - \frac{2E}{m_{3/2}}\right)\; , 
\end{equation}
where
\begin{equation}
D_\gamma = C_\gamma \frac{H_0 \Omega_M^{1/2}}{\Omega_{3/2} \rho_c}
\int_{l.o.s.} \rho_{halo} (\vec{l}) d\vec{l} \; .
\end{equation}
The ratio $ D_\gamma/C_\gamma $ is given only by cosmological
constants and the halo dark matter density integrated along the 
line of sight, so the intensity and angular distribution of the 
halo signal is very sensitive to the distribution of the dark 
matter in the Milky Way. 
It is surprising that for typical halo models, such number
is of order unity~\cite{Asaka:1998ju} and shows moderate 
angular dependence if one excludes the galactic centre region.
The anisotropic part of the halo signal may be partially hidden
in the diffuse galactic $\gamma$-ray emission due to conventional 
astrophysical processes.
We expect therefore the isotropic signal in the extragalactic 
$\gamma$-ray flux to be a combination of both the continuum 
spectrum in Eq.~(\ref{photon-flux}) and part of the halo line 
in Eq.~(\ref{photon-flux-halo}).

Assuming that one understands the diffuse galactic $\gamma$-ray flux, 
one can extract from the EGRET data the extragalactic diffuse
component. 
The first analysis of Sreekumar {\it et al.}~\cite{egret} gave an
extragalactic flux described by the power law
\begin{equation}
E^2 {dJ\over dE} = 1.37 \times 10^{-6}\ 
\left(\frac{E}{1~{\rm GeV}}\right)^{-0.1} 
({\rm cm}^2 {\rm str}~{\rm s})^{-1}{\rm GeV}  
\end{equation}
in the energy range 50 MeV--10 GeV. 
A non-observation of a $\gamma$-ray line can then be used to constrain 
the allowed gravitino mass and lifetime \cite{ty00}. Assuming the
gravitinos to make up all the Cold Dark Matter density, and taking 
a $3\sigma$ upper bound on the flux above $100$ MeV corresponding to 
$2.23 \times 10^{-6} 
({\rm cm}^2 {\rm str}~{\rm s})^{-1}{\rm GeV} $~\cite{egret},
we can have directly a rough bound on the gravitino lifetime 
from $C_\gamma $ as
\begin{equation}
\label{bound-tau}
\tau_{3/2} \gsim 4 \times 10^{26} \mbox{s}\; .
\end{equation}

The more recent analysis of the EGRET data \cite{smr05} shows in the 
50~MeV -- 2~GeV range a power law behaviour, but a clear excess 
between 2~GeV and 10~GeV. The maximal flux allowed by the data
taking into account the model dependence and systematic errors is not
very far from the one obtained in the old analysis, in fact
the integrated flux between 0.1-10 GeV is given as
$(11.1\pm 0.1)\times 10^{-6}{\rm cm}^{-2}~{\rm str}^{-1}~{\rm s}^{-1}$
compared to 
$(14.5 \pm 0.5)\times 10^{-6}
{\rm cm}^{-2}~{\rm str}^{-1}~{\rm s}^{-1}$~\cite{smr05}.

This is precisely the energy range where, based on our lower bound
on the gravitino mass of 5~GeV, one may expect a gravitino signal. 
It is very remarkable that also the measured flux corresponds to 
the expectation of the model for R-parity and B-L breaking discussed 
in Sec.~2 as can be seen from the bound Eq.~(\ref{bound-tau}).
On the other hand we would expect also an anisotropic flux from the
halo component that EGRET does not resolve probably due to the 
galactic background, which is an order of magnitude larger than 
the extracted extragalactic signal. 

The excess in the extragalactic $\gamma$-ray flux above 2 GeV 
from the EGRET data \cite{smr05} has also been related to the 
annihilation of heavy neutralinos in the galactic halo \cite{em04}. 
Due to the current limitations in the determination of the diffuse 
galactic $\gamma$-ray emission \cite{mxx06} theoretical interpretations 
of the EGRET excess remain uncertain at present.  
Clarification can be expected from the 
Gamma Ray Large Area Space Telescope (GLAST)~\cite{GLAST}, 
to be launched this fall, that aims to improve by a 
factor 30--50 the sensitivity of the EGRET satellite 
to the diffuse gamma ray flux in the range  
\mbox{20 MeV--10~GeV}.

Another constraint comes from the neutrino flux.
In the energy range of interest, from about 1 GeV to 1 TeV,
the extraterrestrial neutrino flux is constrained 
by the flux of upward-going muons measured
by the IMB experiment, that does not show any
discrepancy with respect to the expected neutrino flux
from cosmic ray interactions in the atmosphere.
The requirement that the neutrino flux
from gravitino decay does not exceed the observed
flux, translates into a lower bound on the gravitino
lifetime, which is roughly $\tau_{3/2}\gsim 6\times 10^{24}$ s
for $m_{3/2}=1$ TeV, and becomes weaker for smaller 
gravitino masses \cite{ggs91}. This bound is clearly consistent
with a signal in the EGRET data, as discussed above.

\subsection{Collider Signatures}

The collider signatures depend on
the nature of the NLSP. Here we consider the cases
that the NLSP is the lightest stau or the lightest neutralino.

The lightest stau, that we assume mainly right-handed, decays through
${\widetilde \tau_R}\rightarrow \tau \;\nu_\mu, \mu\;\nu_\tau$. 
On the other hand, the small left-handed component of the 
stau mass eigenstate can trigger a decay into two jets
through ${\widetilde \tau_L}\rightarrow b^c t$,
provided the process is kinematically open.
The hadronic decays are enhanced
compared to the leptonic decays by the larger bottom
Yukawa coupling and by the colour factor, but are
usually suppressed by the small left-right mixing.

If the leptonic decay channel is the dominant mode, the decay length 
can be approximated by
\begin{equation}
c \tau^{lep}_{\tilde\tau} \sim
30~{\rm cm}\left(  \frac{m_{\tilde{\tau}}}{200 {\rm GeV}} \right)^{-1}
\left(\frac{\epsilon_2}{10^{-7}}\right)^{-2}
\left(\frac{\tan\beta}{10}\right)^{-2} \;.
\end{equation}
It is intriguing that the sufficient condition 
to avoid the erasure of the baryon asymmetry, Eq.~(\ref{washout}), 
implies the observation of a displaced stau vertex at
future colliders, more than 3mm away from the beam axis for 
$\epsilon_2 < 10^{-6}$.
In the particular case of the flavour model (II) discussed
in Sec.~2.2, $\epsilon_2\sim 6\times 10^{-8}$, 
one has a spectacular signal consisting on a 
heavily ionising charged track of length $\sim 0.8$ m, 
followed by a muon track or a jet and missing energy,
corresponding to ${\widetilde \tau}\rightarrow \mu \nu_\tau$
or ${\widetilde \tau}\rightarrow \tau \nu_\mu$, respectively.

If the hadronic channel ${\widetilde \tau_L}\rightarrow b^c t$ is the 
dominant mode, the decay length is given by
\begin{equation}
c \tau^{ had}_{\tilde\tau}\sim
1.4~{\rm m}\left(\frac{m_{\tilde{\tau}}}{200 {\rm GeV}} \right)^{-1}
\left(\frac{\epsilon_3}{10^{-7}}\right)^{-2}
\left(\frac{\tan\beta}{10}\right)^{-2} 
\left(\frac{\cos\theta_\tau}{0.1}\right)^{-2}\;,
\end{equation}
where $\theta_\tau$ denotes the mixing angle of the staus.
This channel also yields a very unique signature at
colliders, consisting of a heavily ionising charged track
followed by two jets.

These characteristic signatures would allow to 
distinguish at colliders our scenario from the case
with conserved R-parity where the decay $\tilde{\tau}\rightarrow \psi_{3/2}\tau$  leads to (cf.~\cite{bhx04})
\begin{equation}
c\tau^{3/2}_{\tilde{\tau}} \sim 40~{\rm cm} 
\left({m_{3/2}\over 1~{\rm keV}}\right)^2
\left({m_{\tilde{\tau}}\over 200~{\rm GeV}}\right)^{-5}\;.
\end{equation}
Hence, for a gravitino mass $m_{3/2}\lsim \mathcal{O}(10~{\rm keV})$,
the decay length of the lightest stau is shorter than $\mathcal{O}(10~{\rm m})$, and would therefore decay inside the detector
into  tau and  gravitino. The experimental signature 
for this process would be identical to the decay 
$\widetilde \tau\rightarrow \tau \nu_\mu$. However,
the scenario with R-parity violation also predicts
the decay $\widetilde \tau\rightarrow \mu \nu_\tau$, 
with very similar branching ratio due to $SU(2)$ invariance. 
Although this signature
could be mimicked by a scenario with conserved R-parity
if lepton flavour is violated, through the decay
$\widetilde \tau\rightarrow \mu \psi_{3/2}$, large branching
ratios are precluded from present bounds on 
flavour violation \cite{hi04}. In consequence, the observation
of a comparable number of tau and muon events in stau decays
would constitute a signature for the scenario with 
R-parity violation. Also, the observation of a stau decaying
into two jets would undoubtedly point to the scenario
with R-parity violation.

On the other hand, if the lightest neutralino is the NLSP, it 
decays through $\chi^0_1\rightarrow \tau^{\pm} W^{\mp}$ \cite{mrv98}, 
or through $\chi^0_1\rightarrow b\, b^c\, \nu$ \cite{dr91} if the former 
decay channel is kinematically closed. The corresponding decay
lengths can be approximated by
\begin{eqnarray}
&&c \tau^{\rm 2-body}_{\chi^0_1}\sim 20~{\rm cm}
\left(\frac{m_{\chi^0_1}}{200~{\rm GeV}} \right)^{-3}
\left(\frac{\epsilon_3}{10^{-7}}\right)^{-2}
\left(\frac{\tan\beta}{10}\right)^{2} \;, \\ 
&&c \tau^{\rm 3-body}_{\chi^0_1}
 \sim 600~{\rm m}
\left(\frac{m_{\widetilde \nu_L}}{300~{\rm GeV}} \right)^{4}
\left(\frac{m_{\chi^0_1}}{200~{\rm GeV}} \right)^{-5}
\left(\frac{\epsilon_3}{10^{-7}}\right)^{-2}
\left(\frac{\tan\beta}{10}\right)^{-2} \;.
\end{eqnarray}
 
Again, this scenario can be easily discriminated
at future colliders from the scenario with conserved R-parity.
In this case, the neutralino decays into gravitino and 
photon \cite{ddrt96} with decay length
\begin{equation}
c\tau^{3/2}_{\chi^0_1} \sim 80~{\rm cm} 
\left({m_{3/2}\over 1~{\rm keV}}\right)^2
\left({m_{\chi^0_1}\over 200~{\rm GeV}}\right)^{-5}\;.
\end{equation}
For a gravitino mass $m_{3/2}\lsim \mathcal{O}(10~{\rm keV})$
the neutralino would decay inside the detector producing
an energetic photon and missing energy, which is clearly
distinguishable from the signals in the R-parity violating
scenario that in general involve jets.

\subsection{Microscopic Determination of the Planck Mass}

Recently, a method has been proposed for the microscopic determination
of the Planck mass in collider experiments \cite{bhx04}, providing a
direct test of supergravity. The method requires a very long lived stau 
NLSP which decays mostly into tau and gravitino, which is difficult to
reconcile with recent constraints from BBN \cite{pos06,kkm05},
unless there is a late--time entropy production \cite{BHIY}.
In the picture proposed in this letter,
where primordial nucleosynthesis, thermal leptogenesis
and dark matter are naturally consistent, this 
method cannot be pursued, as the stau decays
predominantly in the R-parity violating channel
into charged lepton and neutrino. 

Nevertheless, from a gravitino signal in the diffuse $\gamma$-ray flux
and the width for the stau decay into two jets, one can still obtain a
microscopic estimate of the Planck mass.
The gravitino mass is given by the maximal energy of the photon, 
$m_{3/2}=2 E_{\gamma}$, and the gravitino lifetime can be determined
from the photon flux, Eqs.~(\ref{photon-flux},\ref{photon-flux-halo}). 
Then, using the expression for the gravitino decay rate,
Eq.~(\ref{gamma-gravitino}), one can rewrite the
Planck mass in terms of the gravitino mass, lifetime and
photino-neutrino mixing as 
\begin{eqnarray}
\label{MP}
\Mp & = & \left(\frac{m_{3/2}^3\tau_{3/2}}{32\pi }\right)^{1/2} 
|U_{\tilde\gamma\nu}| \nonumber\\
& = &
2.5\times 10^{18}~{\rm GeV} 
\left(\frac{m_{3/2}}{10~{\rm GeV}}\right)^{3/2}
\left(\frac{\tau_{3/2}}{4\times 10^{27}~{\rm s}}\right)^{1/2}
\left({|U_{\tilde\gamma\nu}|\over 10^{-8}}\right) 
\end{eqnarray}
where $|U_{\tilde\gamma\nu}|$ is related to the 
decay rate of the stau into two jets\footnote{Note that
the decay rate of the stau into leptons depends
on $\epsilon_2$ whereas  $|U_{\tilde\gamma\nu}|$
depends on the sneutrino VEV and therefore mainly
on $\epsilon_3$ for the hierarchical case.}. 
We can cast the dependence on the decay rate as a 
dependence on the decay length of the stau in this channel,
yielding
\begin{equation}
|U_{\tilde\gamma\nu}|^2  
\simeq 10^{-16} 
\left(\frac{c \tau^{had}_{\tilde\tau}}{1.4{\rm m}}\right)^{-1}
\left(\frac{\mt}{200{\rm GeV}}\right)^{-3}
\left(\frac{\tan\beta}{10}\right)^2
\left(\frac{\cos\theta_\tau}{0.1}\right)^{2} \;.
\end{equation}

The measurement of the decay length of the stau
in the hadronic channel, complemented with additional information
about supersymmetry breaking parameters, can provide
a determination of $|U_{\tilde\gamma\nu}|$. 
The measurement of the photon energy and the photon
flux in the diffuse $\gamma$-ray background
then gives the gravitino mass and lifetime and, 
using Eq.~(\ref{MP}), an estimate of the Planck mass.

\section{Conclusions}

On theoretical grounds, theories with and without R-parity are on equal
footing.
In this paper we have presented a simple model where R-parity
is not conserved and its violation is connected to the
scale of B-L breaking. 
One can then have R-parity violating couplings that are small enough
to be consistent with baryogenesis and gravitino dark matter, yet large
enough to allow for the NLSP decay before nucleosynthesis. For gravitino
masses above $5~{\rm GeV}$ one obtains a cosmological history consistent
with thermal leptogenesis, thermally produced gravitino dark matter and
primordial nucleosynthesis.

Relic gravitino decays into neutrino and photon yield a diffuse halo
and extragalactic $\gamma$-ray flux which depends on the R-parity
violating Yukawa couplings. It is remarkable that for a gravitino mass 
$m_{3/2} = {\cal O}(10)$~GeV, the 
predicted photon flux could be part of the apparent excess in the
extragalactic diffuse $\gamma$-ray flux obtained from the EGRET data.
However, given the current uncertainties in the determination of the 
diffuse galactic $\gamma$-ray emission, this consistency may be 
accidental.
Unequivocal evidence for decaying gravitino dark matter could come from 
the results of GLAST.

The flavour dependent pattern of R-parity breaking can give striking 
signatures at the LHC, in particular a vertex of the NLSP, that is
significantly displaced from the beam axis. Together with the
measurement of supersymmetry breaking parameters at the LHC, 
the observation of a redshifted photon spectral line from gravitino 
decay by GLAST can allow a microscopic determination of the Planck mass.
In the less optimistic case where the R-parity breaking Yukawa couplings
are near to their lower bound, astrophysical detection will be very 
challenging whereas signals hinting at R-parity breaking and gravitino 
dark matter could still come from stau decays, as in the case of 
R-parity  conservation.

\section*{Acknowledgements}

We are grateful to G.~Heinzelmann and J.~Ripken for valuable discussions
on the diffuse $\gamma$-ray background, and to Satoshi Shirai.
L. C. acknowledges the support of the ``Impuls- und Vernetzungsfond'' 
of the Helmholtz Association, contract number VH-NG-006.


\begin{thebibliography}{99}

\bibitem{fay78}
  P.~Fayet, {\it Supersymmetric Theories of Particles},
  Orbis Scientiae 1978:0413 (QCD161:C6:1978)

\bibitem{syw81}
  N.~Sakai and T.~Yanagida,
  Nucl.\ Phys.\ B {\bf 197} (1982) 533;\\
  S.~Weinberg,
  Phys.\ Rev.\ D {\bf 26} (1982) 287.

\bibitem{hs84}
  L.~J.~Hall and M.~Suzuki,
  Nucl.\ Phys.\ B {\bf 231} (1984) 419.

\bibitem{axx04}
For recent discussions and references, see\\
  B.~C.~Allanach, A.~Dedes and H.~K.~Dreiner,
  Phys.\ Rev.\ D {\bf 69} (2004) 115002
  [Erratum-ibid.\ D {\bf 72} (2005) 079902],
  [arXiv:hep-ph/0309196];\\
  R.~Barbier {\it et al.},
  Phys.\ Rept.\  {\bf 420} (2005) 1
  [arXiv:hep-ph/0406039].

\bibitem{cxx91}
  B.~A.~Campbell, S.~Davidson, J.~R.~Ellis, K.~A.~Olive,
  Phys.\ Lett.\ B {\bf 256} (1991) 457;\\
  W.~Fischler, G.~F.~Giudice, R.~G.~Leigh and S.~Paban,
  Phys.\ Lett.\ B {\bf 258} (1991) 45;\\
  H.~K.~Dreiner and G.~G.~Ross,
  Nucl.\ Phys.\ B {\bf 410} (1993) 188
  [arXiv:hep-ph/9207221].

\bibitem{ty00}
  F.~Takayama and M.~Yamaguchi,
  Phys.\ Lett.\ B {\bf 485} (2000) 388
  [arXiv:hep-ph/0005214].

\bibitem{pos06}
  M.~Pospelov,
  arXiv:hep-ph/0605215;\\
  R.~H.~Cyburt et al.,
  JCAP {\bf 0611} (2006) 014
  [arXiv:astro-ph/0608562].

\bibitem{kt06}
  K.~Kohri and F.~Takayama,
  arXiv:hep-ph/0605243;\\
  M.~Kaplinghat and A.~Rajaraman,
  Phys.\ Rev.\ D {\bf 74} (2006) 103004
  [arXiv:astro-ph/0606209].

\bibitem{kkm05}
  M.~Kawasaki, K.~Kohri and T.~Moroi,
  Phys.\ Rev.\ D {\bf 71} (2005) 083502
  [arXiv:astro-ph/0408426];\\
  J.~L.~Feng, S.~Su and F.~Takayama,
  Phys.\ Rev.\ D {\bf 70} (2004) 075019
  [arXiv:hep-ph/0404231];\\
  F.~D.~Steffen,
  JCAP {\bf 0609} (2006) 001
  [arXiv:hep-ph/0605306].

\bibitem{BHIY}
  W.~Buchm\"uller, K.~Hamaguchi, M.~Ibe and T.~T.~Yanagida,
  arXiv:hep-ph/0605164.

\bibitem{kkkm06}
  T.~Kanzaki, M.~Kawasaki, K.~Kohri and T.~Moroi,
  Phys.\ Rev.\  D {\bf 75} (2007) 025011
  [arXiv:hep-ph/0609246].

\bibitem{fy86} 
M.~Fukugita and T.~Yanagida, 
Phys.~Lett.~{\bf B174}, 45 (1986).

\bibitem{di02}
  S.~Davidson and A.~Ibarra,
  Phys.\ Lett.\ B {\bf 535} (2002) 25
  [arXiv:hep-ph/0202239].

\bibitem{bdp05}
  W.~Buchm\"uller, P.~Di Bari and M.~Pl\"umacher,
  Annals Phys.\  {\bf 315} (2005) 305
  [arXiv:hep-ph/0401240].

\bibitem{bbb01}
  M.~Bolz, A.~Brandenburg and W.~Buchm\"uller,
  Nucl.\ Phys.\ B {\bf 606} (2001) 518
  [arXiv:hep-ph/0012052].

\bibitem{ps06}
  J.~Pradler and F.~D.~Steffen,
  Phys.\ Rev.\  D {\bf 75} (2007) 023509
  [arXiv:hep-ph/0608344].

\bibitem{rs07}
  V.~S.~Rychkov and A.~Strumia,
  arXiv:hep-ph/0701104.

\bibitem{abc03}
  T.~Asaka, W.~Buchm\"uller and L.~Covi,
  Phys.\ Lett.\ B {\bf 563} (2003) 209
  [arXiv:hep-ph/0304142].

\bibitem{tw06}
  R.~Tatar and T.~Watari,
  Nucl.\ Phys.\ B {\bf 747} (2006) 212
  [arXiv:hep-th/0602238];
  arXiv:hep-ph/0605315.

\bibitem{bhx06}
  W.~Buchm\"uller, K.~Hamaguchi, O.~Lebedev and M.~Ratz,
  arXiv:hep-th/0606187.

\bibitem{gm88}
  G.~F.~Giudice and A.~Masiero,
  Phys.\ Lett.\ B {\bf 206} (1988) 480.

\bibitem{by99}
  W.~Buchm\"uller and T.~Yanagida,
  Phys.\ Lett.\ B {\bf 445} (1999) 399
  [arXiv:hep-ph/9810308].

\bibitem{bilinear}
  S.~Roy and B.~Mukhopadhyaya,
  Phys.\ Rev.\ D {\bf 55} (1997) 7020
  [arXiv:hep-ph/9612447];\\
  M.~Hirsch, M.~A.~Diaz, W.~Porod, J.~C.~Romao and J.~W.~F.~Valle,
  Phys.\ Rev.\ D {\bf 62} (2000) 113008
  [Erratum-ibid.\ D {\bf 65} (2002) 119901],
  [arXiv:hep-ph/0004115];\\
  A.~Abada, S.~Davidson and M.~Losada,
  Phys.\ Rev.\ D {\bf 65} (2002) 075010
  [arXiv:hep-ph/0111332];\\
  E.~J.~Chun, D.~W.~Jung and J.~D.~Park,
  Phys.\ Lett.\ B {\bf 557} (2003) 233
  [arXiv:hep-ph/0211310].

\bibitem{ty99}
  F.~Takayama and M.~Yamaguchi,
  Phys.\ Lett.\ B {\bf 476} (2000) 116
  [arXiv:hep-ph/9910320].

\bibitem{mc01}
  G.~Moreau and M.~Chemtob,
  Phys.\ Rev.\ D {\bf 65} (2002) 024033
  [arXiv:hep-ph/0107286].

\bibitem{PDG}
W.-M. Yao {\it et al.},  
Journal of Physics G 33, 1 (2006).

\bibitem{Asaka:1998ju}
  T.~Asaka, J.~Hashiba, M.~Kawasaki and T.~Yanagida,
  Phys.\ Rev.\  D {\bf 58} (1998) 023507
  [arXiv:hep-ph/9802271].

\bibitem{egret}
  P.~Sreekumar {\it et al.}  [EGRET Collaboration],
  Astrophys.\ J.\  {\bf 494} (1998) 523
  [arXiv:astro-ph/9709257].

\bibitem{smr05}
  A.~W.~Strong, I.~V.~Moskalenko and O.~Reimer,
  Astrophys.\ J.\  {\bf 613} (2004) 962
  [arXiv:astro-ph/0406254];
  Astrophys.\ J.\  {\bf 613} (2004) 956.
  [arXiv:astro-ph/0405441].

\bibitem{em04}
  D.~Els\"asser and K.~Mannheim,
  Phys.\ Rev.\ Lett.\  {\bf 94} (2005) 171302
  [arXiv:astro-ph/0405235].


\bibitem{mxx06}
I.~V.~Moskalenko, S.~W.~Digel, T.~A.~Porter, O.~Reimer and A.~W.~Strong,
arXiv:astro-ph/0609768.

\bibitem{GLAST}
http://www.glast.gsfc.nasa.gov.

\bibitem{ggs91}
  P.~Gondolo, G.~Gelmini and S.~Sarkar,
  Nucl.\ Phys.\ B {\bf 392} (1993) 111
  [arXiv:hep-ph/9209236].

\bibitem{bhx04}
  W.~Buchm\"uller, K.~Hamaguchi, M.~Ratz, T.~Yanagida,
  \mbox{Phys.\ Lett.\ B {\bf 588} (2004) 90}
  [arXiv:hep-ph/0402179].

\bibitem{hi04}
  K.~Hamaguchi and A.~Ibarra,
  JHEP {\bf 0502} (2005) 028
  [arXiv:hep-ph/0412229].

\bibitem{mrv98}
  B.~Mukhopadhyaya, S.~Roy and F.~Vissani,
  Phys.\ Lett.\ B {\bf 443} (1998) 191
  [arXiv:hep-ph/9808265];\\
  E.~J.~Chun and J.~S.~Lee,
  Phys.\ Rev.\ D {\bf 60}, 075006 (1999)
  [arXiv:hep-ph/9811201].

\bibitem{dr91}
  H.~K.~Dreiner and G.~G.~Ross,
  Nucl.\ Phys.\ B {\bf 365}, 597 (1991).

\bibitem{ddrt96}
  S.~Dimopoulos, M.~Dine, S.~Raby and S.~D.~Thomas,
  Phys.\ Rev.\ Lett.\  {\bf 76} (1996) 3494
  [arXiv:hep-ph/9601367].

\end{thebibliography}
\end{document}